%% file: main.tex
\documentclass[sigconf]{acmart}
\AtBeginDocument{%
  }

\setcopyright{acmlicensed}
\copyrightyear{2018}
\acmYear{2018}
\acmDOI{XXXXXXX.XXXXXXX}
\acmConference[Conference acronym 'XX]{Make sure to enter the correct
  conference title from your rights confirmation email}{June 03--05,
  2018}{Woodstock, NY}
\acmISBN{978-1-4503-XXXX-X/2018/06}

\usepackage{subcaption}
\usepackage{tabularx}

\begin{document}

\title{PLUM: Adapting Pre-trained Language Models for Industrial-scale Generative Recommendations}



\settopmatter{authorsperrow=5}

\author{Ruining He}
\authornote{Core contributors. Alphabetical order.} 
\authornote{Corresponding authors: Xinyang Yi \texttt{(xinyang@google.com)}, Lukasz Heldt \texttt{(heldt@google.com)}, Raghunandan Keshavan \texttt{(hkraghunandan@google.com)}, Ruining He \texttt{(ruininghe@google.com)}, Nikhil Mehta \texttt{(nikhilmehta@google.com)}, Alicia Tsai \texttt{(aliciatsai@google.com)}, and Lichan Hong \texttt{(lichan@google.com)}.}
\affiliation{%
  \institution{Google DeepMind}
  \country{}
}

\author{Lukasz Heldt}
\authornotemark[1] 
\authornotemark[2]
\affiliation{%
  \institution{YouTube}
  \country{}
}

\author{Lichan Hong}
\authornotemark[1]
\authornotemark[2]
\affiliation{%
  \institution{Google DeepMind}
  \country{}
}

\author{Raghunandan Keshavan}
\authornotemark[1]
\authornotemark[2]
\affiliation{%
  \institution{YouTube}
  \country{}
}

\author{Shifan Mao}
\authornotemark[1]
\affiliation{%
  \institution{YouTube}
  \country{}
}

\author{Nikhil Mehta}
\authornotemark[1]
\authornotemark[2]
\affiliation{%
  \institution{Google DeepMind}
  \country{}
}

\author{Zhengyang Su}
\authornotemark[1]
\affiliation{%
  \institution{YouTube}
  \country{}
}

\author{Alicia Tsai}
\authornotemark[1]
\authornotemark[2]
\affiliation{%
  \institution{Google DeepMind}
  \country{}
}

\author{Yueqi Wang}
\authornotemark[1]
\affiliation{%
  \institution{YouTube}
  \country{}
}

\author{Shao-Chuan Wang}
\authornotemark[1]
\affiliation{%
  \institution{Google DeepMind}
  \country{}
}

\author{Xinyang Yi}
\authornotemark[1]
\authornotemark[2]
\affiliation{%
  \institution{Google DeepMind}
  \country{}
}

\author{Lexi Baugher}
\affiliation{%
  \institution{YouTube}
  \country{}
}

\author{Baykal Cakici}
\affiliation{%
  \institution{YouTube}
  \country{}
}

\author{Ed Chi}
\affiliation{%
  \institution{Google DeepMind}
  \country{}
}

\author{Cristos Goodrow}
\affiliation{%
  \institution{YouTube}
  \country{}
}

\author{Ningren Han}
\affiliation{%
  \institution{YouTube}
  \country{}
}

\author{He Ma}
\affiliation{%
  \institution{YouTube}
  \country{}
}

\author{Romer Rosales}
\affiliation{%
  \institution{YouTube}
  \country{}
}

\author{Abby Van Soest}
\affiliation{%
  \institution{YouTube}
  \country{}
}

\author{Devansh Tandon}
\affiliation{%
  \institution{YouTube}
  \country{}
}

\author{Su-Lin Wu}
\affiliation{%
  \institution{YouTube}
  \country{}
}

\author{Weilong Yang}
\affiliation{%
  \institution{YouTube}
  \country{}
}

\author{Yilin Zheng}
\affiliation{%
  \institution{YouTube}
  \country{}
}

\renewcommand{\shortauthors}{He et al.}

\begin{abstract}
\input{abstract}
\end{abstract}




\maketitle

\section{Introduction}
\input{introduction}

\section{PLUM Framework}
\input{plum_framework}

\section{Experiments}
\input{experiments}

\section{Conclusion and Future Work}
\input{conclusion}

\section{Acknowledgements}
\input{ack}

\bibliographystyle{ACM-Reference-Format}
\bibliography{references}


\appendix
\section{Appendix}
\input{appendix}

\end{document}

%% file: abstract.tex
  Large Language Models (LLMs) pose a new paradigm of modeling and computation for information tasks. Recommendation systems are a critical application domain poised to benefit significantly from the sequence modeling capabilities and world knowledge inherent in these large models. In this paper, we introduce PLUM, a framework designed to adapt pre-trained LLMs for industry-scale recommendation tasks. PLUM consists of item tokenization using Semantic IDs \citep{rajput2023genrecs, singh2024sid}, continued pre-training (CPT) on domain-specific data, and task-specific fine-tuning for recommendation objectives. For fine-tuning, we focus particularly on generative retrieval, where the model is directly trained to generate Semantic IDs of recommended items based on user context. We conduct comprehensive experiments on large-scale internal video recommendation datasets. Our results demonstrate that PLUM achieves substantial improvements for retrieval compared to a heavily-optimized production model built with large embedding tables. We also present a scaling study for the model's retrieval performance, our learnings about CPT, a few enhancements to Semantic IDs, along with an overview of the training and inference methods that enable launching this framework to billions of users in YouTube.

%% file: introduction.tex
Recommender systems are critical in modern digital platforms, profoundly shaping how users discover and interact with content. The last decades have witnessed the remarkable success of deep learning models for recommendation in both retrieval and ranking stages. Despite the use of neural networks, the dominant paradigm in industrial recommenders continues to rely on massive embedding tables to represent high-cardinality categorical features such as item IDs (see, e.g., \citep{Covington2016, naumov2019dlrm, liu2022monolith, Coleman2023}). In these Large Embedding Models (LEMs), the vast majority of parameters reside within these embedding tables. While highly effective at memorizing user-item interactions, this architectural choice hinders the potential benefits of deeper, more complex networks (e.g., \citep{chen2019behaviorseq, gu2020dit}). This scaling methodology, focused on enlarging embedding tables, is in contrast to that of LLMs, which emphasizes on increasing neural network sizes to learn compositions of compact input tokens.

The success of LLMs has inspired an emerging paradigm shift for building recommendation models \citep{rajput2023genrecs, pmlr-v235-zhai24a, firooz2025360brew, zhou2025onerectechnicalreport, pang2025gram, chen2025pinfm}. The inherent sequence modeling capabilities and vast world knowledge encoded in LLMs present opportunities for building more intelligent and personalized recommender systems. However, adapting LLMs for recommendations is non-trivial. The primary challenge lies in bridging the domain gap: LLMs are not pre-trained with user behavior data and item corpus in the target domain, making it harder for them to understand user preferences from user activities and nuanced item quality. Consequently, directly applying off-the-shelf LLMs to recommendation tasks has shown a persistent performance gap, even on small public datasets \citep{hou2024llmranker}. Secondly, the traditional input representation based on large embedding tables poses a scaling challenge. These tables demand large amounts of training data, making it expensive to train large Transformer architectures.

This paper takes a step in the direction of teaching LLMs to solve recommendation tasks. We introduce \textbf{PLUM}, a framework for effectively adapting pre-trained LLMs for industrial-scale generative recommendation. PLUM consists of three key stages:
\begin{itemize}
    \item \textbf{Item tokenization}. Each item in the corpus is represented by a sequence of discrete tokens known as Semantic IDs (SIDs). Building on the prior work with RQ-VAE \citep{rajput2023genrecs, singh2024sid}, we introduce a set of new techniques (referred to as SID-v2) for incorporating user behavioral signals and multi-modal content embeddings, and improve the hierarchical integrity through multi-resolution codebooks and progressive masking.
    \item \textbf{Continued pre-training (CPT)}. In this stage, the vocabulary of a pre-trained LLM is expanded to include the new SID tokens. The model is further pre-trained on a mixture of domain-specific item data and user sequences, and general-domain text data to align the new SID modality with model's existing knowledge.
    \item \textbf{Task-specific fine-tuning}. Finally, the model is fine-tuned for specific recommendation objectives. While PLUM can support various downstream tasks, this paper focuses on its application to generative retrieval, where a decoder-only model is trained to autoregressively generate the SIDs of next items a user is likely to engage with. Generative retrieval does not need to maintain a separate index of an item corpus, and bypasses the dot-product limitation \citep{weller2025theoreticallimit} in embedding-based retrieval \citep{Covington2016, yi2019sampling-bias}.
\end{itemize}

The PLUM framework is designed to address the aforementioned challenges. The continued pre-training bridges the domain gap by enriching the pre-trained LLMs with domain corpus and user behavior patterns. We observe that the model after CPT can demonstrate basic few-shot learning capability for generating text tokens based on SID input. The SID-based input representation circumvents the scaling bottleneck of LEMs. Our experiments show that by shifting the model complexity from input embeddings to the neural networks, PLUM-based generative retrieval can achieve significantly better sample efficiency compared to a production Transformer-based retrieval model with large embedding tables. Because of this, while PLUM retrieval uses a Transformer architecture that has 100x dense parameters than LEM, the overall training cost for the retrieval task is comparable due to faster convergence. We also observe effective scaling: the retrieval performance continues to improve up to a Mixture-of-Experts (MoE) model with over 900M activated parameters (approximately 4.2B parameters in total).

In addition to these contributions, we will share practical learnings from deploying PLUM-based retrieval in a large-scale production environment, particularly on the comparisons between PLUM retrieval and LEM-based retrieval in online A/B testing. The PLUM framework is in production for YouTube recommendations, serving the retrieval of both long-form and short-form videos on multiple core surfaces via online and offline inference.

\subsection{Related Work}

We've seen two primary lines of work inspired by the recent advancements of LLMs. The first line of work focuses on scaling up the neural network component of LEMs with  Transformer architecture (or its variant), which we discuss in Section \ref{sec:transformer-based-recommenders}. The second line, as detailed in Section \ref{sec:sid-quantization} and \ref{sec:generative-retrieval}, focuses on token-based generative recommendation, and recasts recommendation retrieval as seq2seq transduction tasks.

\subsubsection{Sequential Recommendations}
\label{sec:transformer-based-recommenders}
Modeling user sequences and feature interactions have been key areas for improving recommendation models, ranging from using classic sequential neural networks \citep{hidasi2016rnn, kang2018sasrec} to designing custom feature interaction components \citep{wang2021dcn, zhang2024wukong}. More recently, the emerging trend is to use a Transformer-like architecture to unify user history and heterogeneous features in a single sequence \citep{pmlr-v235-zhai24a, huang2025generativeranking, han2025MTGR}, showing better scaling compared to traditional MLP architectures. Furthermore, the training
paradigm itself has evolved to mirror that of LLMs, with some works adopting self-supervised pre-training objectives followed by task-specific fine-tuning \citep{netflix2024foundation, chen2025pinfm}. A common thread across all these advancements is their continued reliance on massive embedding tables for categorical features. Our work departs from this paradigm by replacing large embedding tables with compact input tokens. Besides SIDs, we use LLMs to process heterogeneous features by tokenizing numerical features or simply use dense embedding features as soft tokens.

\subsubsection{Semantic IDs and Quantization}
\label{sec:sid-quantization}
The concept of representing items as sequences of discrete tokens, or Semantic IDs (SIDs), has emerged as a powerful alternative to traditional ID embeddings. This approach allows items to be treated
as a "language" that can be processed by sequence models.  Early work demonstrated the feasibility of training retrieval
models from scratch to predict SIDs \citep{rajput2023genrecs}, and subsequent research showed that SIDs could be hashed to replace random item ID embeddings, enhancing generalization of ranking models \citep{singh2024sid, zheng2025enhancingembedding}.

A significant body of recent research has focused on improving the quality and expressiveness of SIDs by incorporating diverse signals into the quantization process. This includes fusing multi-modal content features \citep{xu2025mmq} and injecting collaborative filtering signals from user behaviors \citep{penha2025semanticidsjointgenerative, li2025bbqrec, yao2025saviorrec, ye2025das}. PLUM contributes to this line of research in parallel. Different from MMQ \citep{xu2025mmq}, where different tokens are generated for each modality, we simply concatenate multiple content embeddings before encoding them with an MLP, offering the flexibility of supporting more embeddings. For incorporating user behaviors, while most works fuse collaborative-filtering (CF) item embeddings to content embeddings during quantization, the CF-based item embeddings are usually dynamic depending on item popularity change, necessitating frequent retraining of the quantizer and downstream models. We avoid this by using CF signals in a contrastive training objective to guide the quantizer to capture useful information from content information. In addition, we present two other key innovations for improving the alignment of SID training with generative retrieval.

\subsubsection{Generative Retrieval and Alignment}
\label{sec:generative-retrieval}
Generative retrieval reframes recommendation retrieval as a sequence-to-sequence task where the model directly generates the SIDs of relevant items in an autoregressive manner. This idea has gained significant traction across the industry (e.g., \citep{wang2025generativepoi, mei2025sidmusic, ju2025generativerecommendationsemanticids, pang2025gram}), since the seminal works in document search \citep{Tay2022DSI} and recommendation \citep{rajput2023genrecs}. For instance, OneRec \citep{zhou2025onerectechnicalreport} adopts an encoder-decoder architecture to generate video SIDs, and leverages RL training to enhance the model quality with a reward model. Other works \citep{yang2024LIGER, yang2025sparsemeetsdenseunified} propose hybrid models that generate both SIDs and dense embeddings to alleviate the information loss from quantization.
A commonality among most of these approaches is their focus on training generative models from scratch. Our work has a focus on aligning SIDs with LLMs, and investigates the impact of LLM pre-training and CPT on generative retrieval performance. While some research \citep{cao2024aligninglargelanguagemodels, firooz2025360brew} has explored aligning LLMs with recommendation tasks, their focus is primarily on language interfaces. There is limited study of how to effectively incorporate SIDs as a new modality.

%% file: plum_framework.tex
\subsection{Semantic IDs}

\begin{figure*}[h!]
    \centering
    \includegraphics[width=0.65\textwidth, clip, trim=0.1cm 0.1cm 0.1cm 0.1cm]{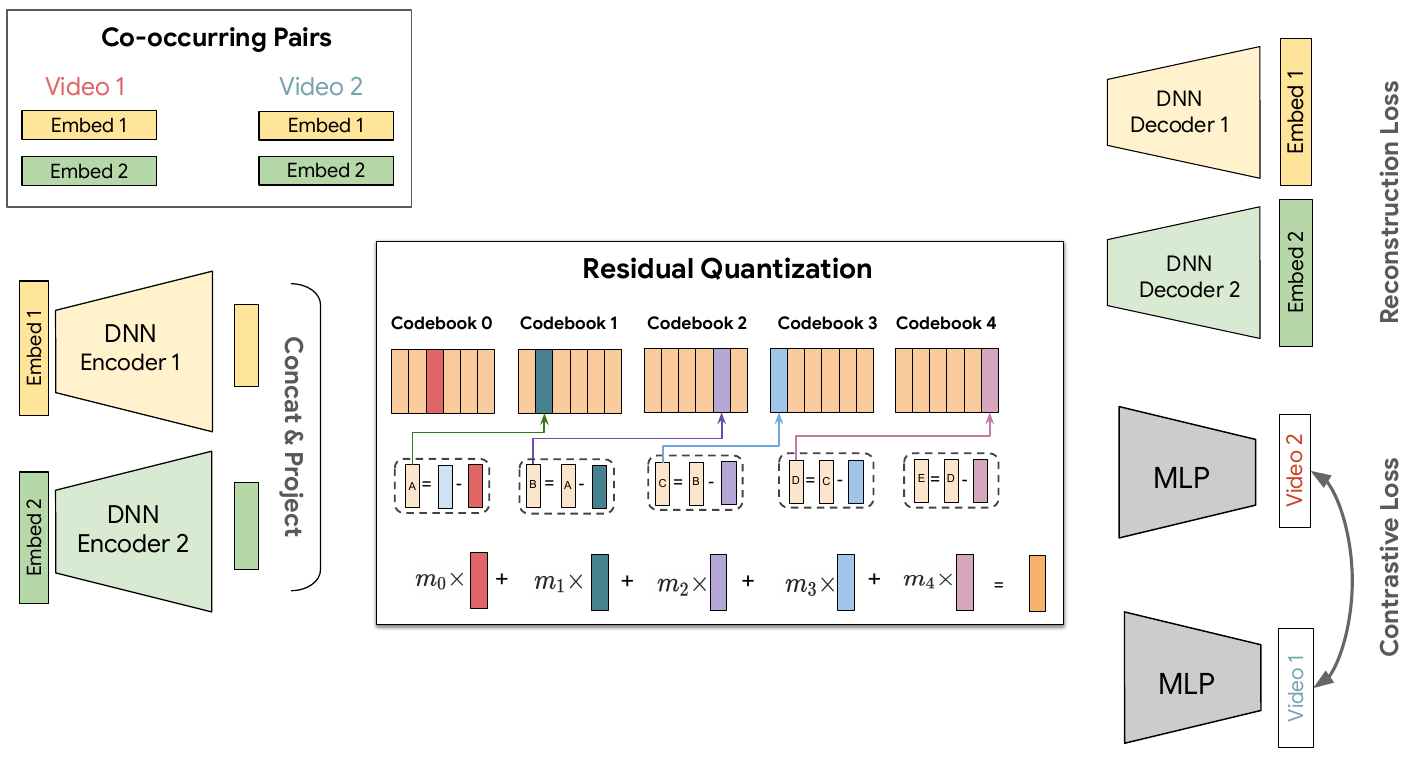}
    \caption{Illustration of our Semantic ID model. It takes two multi-modal video embeddings, encodes them, and compresses the result into a quantized ID using a residual quantizer. This ID is trained to both reconstruct the original inputs and semantically cluster co-occurring videos using a contrastive loss.}
    \label{fig:sidv2_model_illustration}
\end{figure*}

A Semantic ID (SID) is conceptualized as a tuple of discrete codewords derived from an item's underlying content features. This process involves two primary stages: (1) encoding high-level content features into a dense semantic embedding, and (2) quantizing this embedding into a hierarchical tuple of codewords. The efficacy of a generative retrieval model is fundamentally dependent on the semantic richness and structural integrity of the SIDs. Building upon the foundational TIGER framework, which leverages a Residual-Quantized Variational AutoEncoder (RQ-VAE), we introduce a series of significant enhancements to this generation process. The initial approach was limited by its reliance on a single content embedding source and an absence of collaborative signals in the ID structure. Our methodology is designed to produce SIDs that are more comprehensive, better aligned with user behavior, and more hierarchically coherent. Figure \ref{fig:sidv2_model_illustration} illustrates our overall design for training the SID model.

\subsubsection{Fused Multi-Modal Content Representation}
A primary limitation of relying on a single content representation in ~\cite{rajput2023genrecs} is its inherent inability to capture the multi-faceted nature of complex media items. For instance, a video's full semantic meaning is conveyed through a combination of its textual metadata, visual content, and audio. A unimodal representation, by definition, overlooks rich, orthogonal sources of information. To overcome this, we architect our framework to be agnostic to the specific number and type of input sources, enabling it to ingest and fuse multiple, heterogeneous representations. The model is designed to accept a set of distinct embedding vectors $\{x_m\}_{m=1}^M$ for each item. The fusion of such multiple representations is achieved through a separate embedding encoder $\mathcal{E}_m$ that encodes $x_m$ into a latent vector $z_m$ and a concatenation $\tilde{z} = [z_1, \dots, z_M]$ followed by a projection, creating a single, unified feature encoded vector $z$. By integrating information from across modalities, we form a superior input for the subsequent RQ-VAE quantization process, ensuring that the resulting SIDs are grounded in a comprehensive understanding of the item's content.

\subsubsection{Hierarchical Refinements in Quantization}

We further refine the RQ-VAE architecture itself to produce a more efficient and meaningful hierarchy. This involves two key innovations:

\begin{itemize}
\item \textbf{Multi-Resolution Codebooks:} Prior work~\citep{rajput2023genrecs} employs a fixed uniform-resolution codebooks that can be parametrically inefficient, leading to a vast and sparsely populated SID space where the majority of potential codeword combinations remain unassigned. We replace this with a multi-resolution codebook structure, where the initial SID levels have high resolution and are maximally discriminative, while subsequent codewords encode low-entropy residuals with lower resolution. In particular, the codebook cardinality is a function of the quantization level: $2048 / 2^{level - 1}$ resulting in a more compact and efficient SID.

\item \textbf{Progressive Masking:} To enforce a stricter and more interpretable hierarchy during residual quantization, we introduce progressive masking. In particular, we define a binary mask scalar $m_l \in \{0,1\}$ where $l$ is the codebook level such that $m_l = \mathbf{1}_{l<r}$ where $r \in [1, L]$ is a random integer and $L$ is the total number of codebook levels. This mask is applied to select the first $r$ codebook levels in SID training.
\end{itemize}

\subsubsection{RQ-VAE with Co-occurrence Contrastive Regularization}
SIDs derived purely from an item's intrinsic content may not fully capture the notion of similarity as perceived by users in a recommendation context. User behavior provides a powerful, extrinsic signal of video relatedness; videos that are frequently watched together are often semantically linked. To bridge this gap between content-based and behavior-based similarity, we inject a strong collaborative signal directly into the ID generation phase by introducing a co-occurrence contrastive loss term ($\mathcal{L}_{con}$) into the RQ-VAE training objective. This objective function is designed to shape the embedding space based on item co-occurrence patterns within user interaction sequences. The loss encourages the model to generate similar SID representations for items that frequently appear together, while pushing apart the representations of items that do not. In particular, the contrastive loss is defined as: \begin{equation}
    \mathcal{L}_{con} =  - \sum_{i=1}^{2N_b} \frac{exp(sim(p_i, p_i^{+}))}{\sum_{j=1}^{2N_b} exp(sim(p_i, p_j))},
\end{equation}
where $p_i$ represents a video representation in a batch of $N_b$ videos, $p_i^*$ represents the representation of the video that co-occurred with video $i$ and $sim(p_i, p_j)$ is the dot-product similarity between videos $i$ and $j$.

\subsubsection{SID Training Loss}

In addition to the training loss introduced in ~\cite{singh2024sid}, we introduce the co-occurence contrastive loss term. In particular, the overall loss is defined as:
\begin{equation}
    \mathcal{L} = \mathcal{L}_{recon} + \mathcal{L}_{rq} + \mathcal{L}_{con},
\end{equation}
where $\mathcal{L}_{recon} = \sum_{m=1}^M ||x_m - \hat{x}_m||^2$ is the reconstruction loss for each multi-modal embedding and $\mathcal{L}_{rq} = \sum_{l=1}^L \beta ||r_l - sg[e^l_*]||^2 + ||sg[r_l] - e^l_*||^2$, where $e^l_*$ is the closest code in the codebook at level $l$, $r_l = r_{l-1} - e^l_*$ is the residual computed recursively at each level starting with $r_{0} = z$ and sg is the stop-gradient operator. As shown in figure~\ref{fig:sidv2_model_illustration}, the final quantized vector $\hat{z}$ is computed with progressive masking as the sum of chosen codes from each level: $\hat{z} = \sum_{l=1}^L m_l \enspace e^l_*$. The quantized vector is then used as an input for each decoder $\mathcal{D}_m$ to reconstruct the input embeddings $\hat{x}_m$.

\subsection{Continued Pre-training}
Following the creation of the SID vocabulary, one primary objective of the continued pre-training (CPT) stage is to develop a base model checkpoint where the SID tokens are semantically grounded and aligned with existing text tokens of the base language model. To achieve this, we train the model using next-token predictions on a large-scale corpus constructed from two primary sources:
\begin{itemize}
    \item \textbf{User behavior data}: This source is essential for personalization and capturing watch histories. During training, each example leverages user watch histories along with additional watch features to model user behavior sequences.
    \item \textbf{Video metadata corpus}: This large-scale corpus is designed to create a strong association between SIDs and their corresponding textual features. Each example includes a video's SID, title, description, ASR captions, channel name, and synthetically generated data.
\end{itemize}
Table \ref{tab:pretraining_data_schema} illustrates our CPT training data schema.

\begin{table}[h!]
    \footnotesize
    \centering
    \caption{Example schemas used in continued pre-training.}
    \label{tab:pretraining_data_schema}
    \begin{tabularx}{\linewidth}{X}
        \toprule
        \textbf{Example user behavior training data} \\
        \midrule
        wh = \texttt{<sid\_1> <channel\_name> <watch\_ratio> <watch\_time>} \\
        \texttt{<hours\_since\_final\_watch> <sid\_2> <channel\_name>} ... || \texttt{<sid\_n>} \\
        \midrule
        \textbf{SID + video title} \\
        \midrule
        Video \texttt{<sid>} has title (en): \texttt{<video\_title>} \\
        \midrule
        \textbf{SID + video topics} \\
        \midrule
        The topics in video \texttt{<sid>} are: \texttt{<topics>} \\
        \bottomrule
    \end{tabularx}
\end{table}

\paragraph{Data Mixture and Training.}
The training set for the CPT stage is composed of a mixture of the user behavior data and the video metadata corpus, with each source accounting for 50\% of the training examples. The CPT stage undergoes 1 million training steps with a batch size of 16, amounting to approximately 260 billion tokens in total. We set up various evaluations to measure the performance of generating SIDs based on user history, the ability to jointly model SIDs and language through held-out video metadata corpus and tracking the degradation of general language capabilities through standard text benchmarks.

\paragraph{In-context Learning.}
The CPT models, having been trained on a mixed corpus of SIDs and natural language, retain the flexibility to generate free-form text, and have the in-context few-shot learning capability. We provide some examples in Appendix \ref{app:in-context-learning}.

\subsection{Generative Retrieval}

While the continued pre-training (CPT) stage enables the model to generate next-video SIDs from user history, a subsequent Supervised Fine-Tuning (SFT) stage is essential to specialize it for the retrieval task. This SFT stage allows the model to incorporate a richer set of features, particularly real-time context, and to be optimized directly for a reward signal that aligns with user experience.

\begin{figure}[h!]
    \centering
    \includegraphics[width=0.9\linewidth]{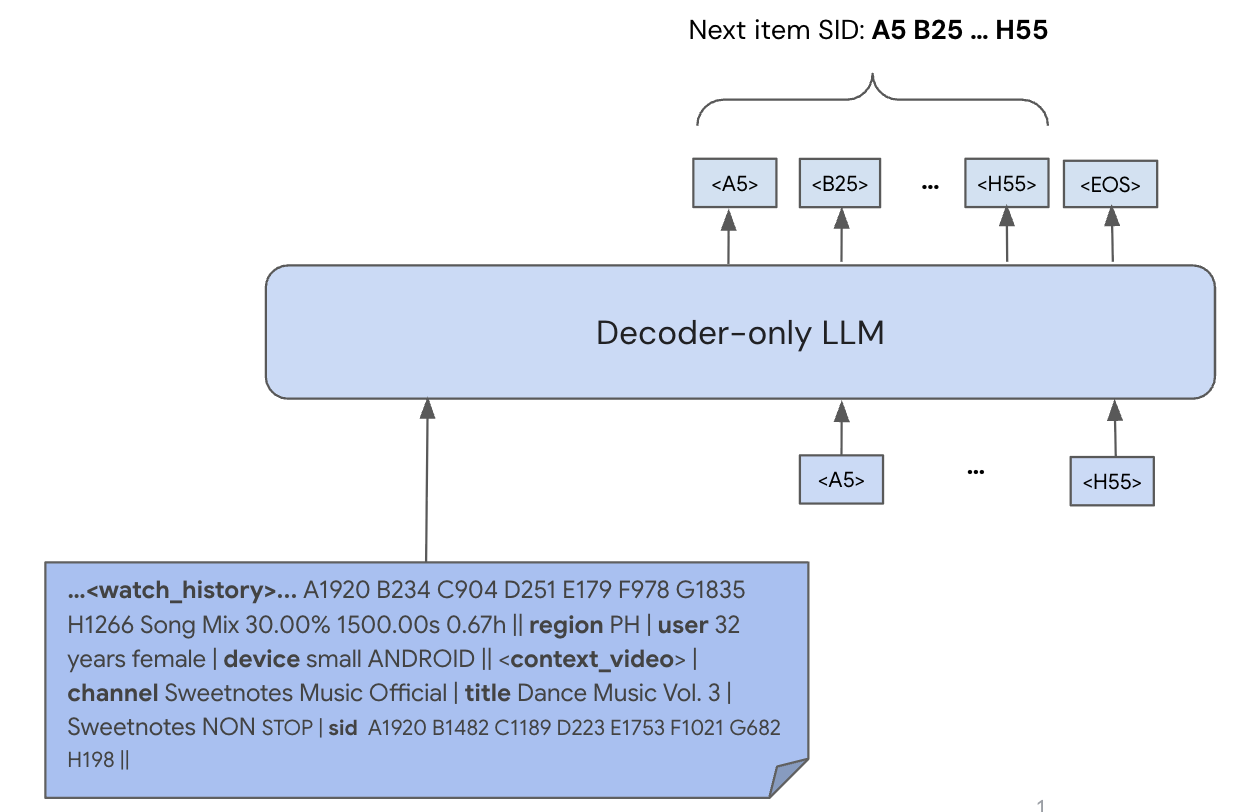}
    \caption{Illustration of Generative Retrieval for next video recommendation. The input prompt is a sequence of interleaved SID tokens, text and custom tokens for numerical features.}
    \label{fig:gen_retrieval_illustration}
\end{figure}

This fine-tuning employs a standard autoregressive, maximum-likelihood objective. The model learns to predict the SID tokens of ground-truth videos, which are defined as clicked videos from user logs given user context and history. Concretely, the model is trained to minimize the following loss:

\begin{equation}
\label{eq:sft_loss}
\mathcal{L}_{\text{SFT}} = - \sum_{t=1}^{L} r(\text{user}, v_{click}) \cdot \log P(sid_t | \text{Context}_{\text{user}}, \text{History}_{\text{user}}, sid_{<t}),
\end{equation}
where $[sid_1,..,sid_L]$ represents the SID of clicked video $v_{click}$, and $r(user, v_{click})$ is a handcrafted reward signal of each click. In practice, given the high cost of training, we sample training examples based on this reward and then weigh the sampled examples equally in the loss. As illustrated in Figure \ref{fig:gen_retrieval_illustration}, the input prompt contains not only SID tokens and custom tokens for numerical features, but also other text features that can be naturally encoded by pre-trained LLMs.

During inference, we use beam search to decode multiple SID sequences, which serve as the set of retrieved candidates. Each generated SID is then mapped to a real video in our billions-scale corpus. While this generative process can potentially produce invalid SIDs (hallucination) or SID-to-video collisions, we observe that the hallucination rate after SFT is very low (< 5$\%$), and the uniqueness of SID-to-video mapping remains high (see Table \ref{tab:sid-ablations}).

%% file: experiments.tex
In this section, we conduct a comprehensive set of experiments to validate the PLUM framework. We first evaluate the performance of a PLUM-based generative retrieval model against a Transformer-based large-embedding retrieval model that contributes to a majority of impressions in production (Section \ref{subsec:perf_gr}). Following this, Section \ref{subsec:sid_ablation} demonstrates the effectiveness of our proposed enhancements to SIDs. We then conduct ablation studies for two critical components in the PLUM framework: Section \ref{subsec:cpt_impact} quantifies the precise impact of the continued pre-training (CPT) stage and the value of initializing from a pre-trained LLM. We finally present a detailed scaling study to analyze the relationship between model size, compute, and retrieval performance (Section \ref{subsec:scaling_study}).

\subsection{Performance of Generative Retrieval}
\label{subsec:perf_gr}

Traditional recommender systems use large embedding models to recommend candidates. However, generative retrieval has a number of advantages over traditional systems. In this section, we study the effectiveness of generative retrieval on the YouTube production setup.


\subsubsection{Experiment Setup}

\paragraph{Model.} For the following experiments, we trained a 900M activated-param PLUM model from the Gemini-1.5 Mixture-of-Experts (MoE) family on both Long Form Video (LFV) and Shorts. The model is warm started from Gemini and continuously finetuned on new engagement data as described above. The training data is a mixture of the most recent data and historical data. During serving, target sequences are decoded using beam search. In our experiments, beam search performed better than random decoding, although at the cost of some diversity.

\paragraph{Baseline.} We compare the performance of the above model to traditional LEM retrieval. This baseline is the top-performing production model, being highly-optimized since early work such as \citet{chen2019}. This model is also based on the Transformer architecture, but most of its parameters are in the embedding layer, with O(10M) vocab sizes for input and output item IDs. Specifically, LEM's neural network comprises only $0.4\%$ of its total parameters, whereas PLUM's neural network accounts for $90\%$.


\subsubsection{Experiment Results}

\paragraph{Recommendation Quality.} 
We assess the quality of recommendations on a few different axes. First, we measure the effective vocab size of a recommender as the number of unique videos needed to cover 95\% of its impressions. Higher vocab sizes are generally desirable for personalized discovery of niche content, showing the model can generalize better. Next, we compare the effectiveness of recommendations using two metrics: a) The click-through-rate (CTR) and Recommendation Acceptance. For Recommendation Acceptance, we compare both the length of video watched (WT/View) and the fraction of video watched (WF/View). In all cases, we report the ratio of the metric for the 900M MoE to the metric for the LEM model.

\begin{table}[ht]
\centering
\caption{Comparison of recommendation quality: Each number is a ratio, dividing the metric for PLUM by that of LEM.}
\label{tab:scaling-configs}
\begin{tabular}{l c c}
\toprule
\textbf{Metric} & \textbf{LFV} & \textbf{Shorts} \\
\midrule
Effective Vocab Size & $2.60\text{x}$ & $13.24\text{x}$ \\
CTR & $1.42\text{x}$ & $1.33\text{x}$ \\
WT/View & $0.72\text{x}$ & $1.13\text{x}$ \\
WF/View & $1.32\text{x}$ & $1.03\text{x}$ \\
\bottomrule
\end{tabular}
\end{table}

We observe that the PLUM model achieves much larger effective vocab size, while having competitive performance against LEM based on the metrics related to user reactions.

\paragraph{Live Experiments.}
We ran live experiments by adding the PLUM model recommendations to the candidate pool. For a fair comparison, we increased the quota (by the same amount) for the best performing retrieval model in production and use this as the baseline. We denote this modified production baseline as LEM+. In the following table, we report the metric movements observed with respect to LEM+ on four key metrics, for both LFV and Shorts. This demonstrates that PLUM-based retrieval can add unique value on top of the existing system, even without large embedding tables.

\begin{table}[ht]
\centering
\caption{Comparison of engagement: Percentage change by adding PLUM compared to LEM+}
\label{tab:scaling-configs}
\begin{tabular}{l c c}
\toprule
\textbf{Metric} & \textbf{LFV} & \textbf{Shorts} \\
\midrule
Engaged Users & $+0.07\%$ & $+0.28\%$ \\
Panel CTR & $+0.76\%$ & $+4.96\%$ \\
Views & $+0.80\%$ & $+0.39\%$ \\
Satisfaction & $+0.06\%$ & $+0.39\%$ \\
\bottomrule
\end{tabular}
\end{table}

\paragraph{Sample Efficiency.}
In our studies, PLUM models are extremely sample efficient. The 900M MoE model trains on roughly 250M examples every day. By comparison, the traditional LEM trains on several billion examples per day. This sample efficiency also makes these models practical for production use, while training cost per example is much larger. Compared to the LEM, the 900M MoE model uses < 0.55x flops to train.

\subsection{Semantic IDs Ablation Study}
\label{subsec:sid_ablation}
In Table~\ref{tab:sid-ablations} we discuss the importance of different changes we introduced to the prior SID~\citep{rajput2023genrecs}. We refer to the prior work as SIDv1, and refer to our proposed method as SIDv2 in the table. Since each SID model will result in a different indexing structure, we report the index uniqueness depicting how many videos in the corpus are uniquely represented in the SID space along with the final Video Recall@10 after training a GenRetrieval model on the SID. The Video Recall@K is determined by calculating the recall of successfully retrieving a specific video given PLUM's top K SID predictions. When an SID maps to multiple videos, we randomly sample a video when computing the recall metric.

We conducted ablation experiments to the SIDv2 changes in a simple setting with a 900M MoE PLUM retrieval model without CPT and removed watch history in the prompt. As shown in Table~\ref{tab:sid-ablations} , these experiments assessed both the uniqueness of Semantic ID and downstream generative retrieval VID Recall@10. Results show that all changes improve generative item retrieval recall. Notably, incorporating a Co-occurrence alignment task greatly improved both Semantic ID uniqueness and recall.

\begin{table}[h!]
\small
\centering
\begin{tabularx}{0.9\linewidth}{X X X }
\toprule
\textbf{SID Model} & \textbf{SID Uniqueness} & \textbf{VID Recall@10} \\
\midrule
SIDv1 (Baseline) & 94.0\% & 12.3\% \\
SIDv2 (Ours) & 96.7\% & 14.4\% \\
\midrule
Ablate Multi-Resolution & 94.8\% & 13.2\% \\
Ablate Multi-Embedding & 96.9\% & 12.8\% \\
Ablate Co-occurence & 91.8\% & 12.6\% \\
\bottomrule
\end{tabularx}
\caption{Ablation experiments on SIDv2 changes}
\label{tab:sid-ablations}
\end{table}


\subsection{Impact of Continued Pre-training}
\label{subsec:cpt_impact}

To measure the importance of the continued pre-training phase, we evaluate its impact on the training efficiency and retrieval performance of the end models for generative retrieval.

\subsubsection{Experiment Setup}
Our training methodology contains two stages: continued pre-training and generative retrieval fine-tuning, each tailored with a specific learning objective and a corresponding set of hyperparameters. To evaluate the distinct contributions of each step, we design a controlled 2x2 ablation study. This involves four model configurations that share the same decoder-only Transformer architecture. We use the MoE-900M architecture for this study. The four models are as follows:
\begin{itemize}
\item \textbf{R1: retrieval SFT (random init).} Transformer model is directly trained on the generative retrieval task with randomly initialized weights.
\item \textbf{R2: retrieval SFT (LLM init).} This model is initialized from a pre-trained LLM checkpoint, but bypasses the CPT stage.
\item \textbf{CR1: CPT (random init) + retrieval SFT.} The model training follows  the two PLUM training stages, but the model used in CPT is randomly initialized.
\item \textbf{CR2: CPT (LLM init) + retrieval SFT.} This represents the full and intended implementation of our proposed framework. 
\end{itemize}

\subsubsection{Experiment Results}
\begin{figure*}
    \centering
    \includegraphics[width=0.49\linewidth]{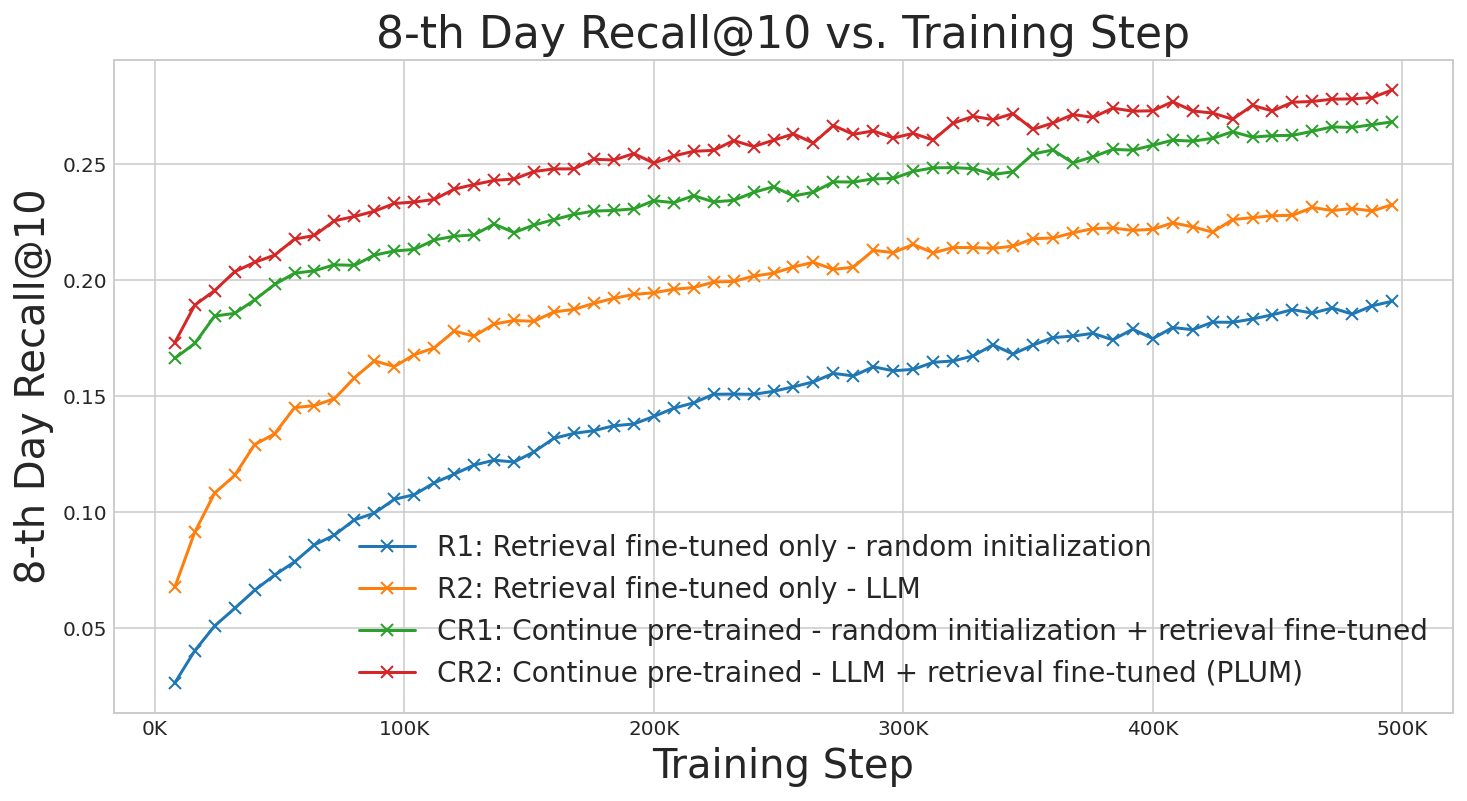}
    \includegraphics[width=0.49\linewidth]{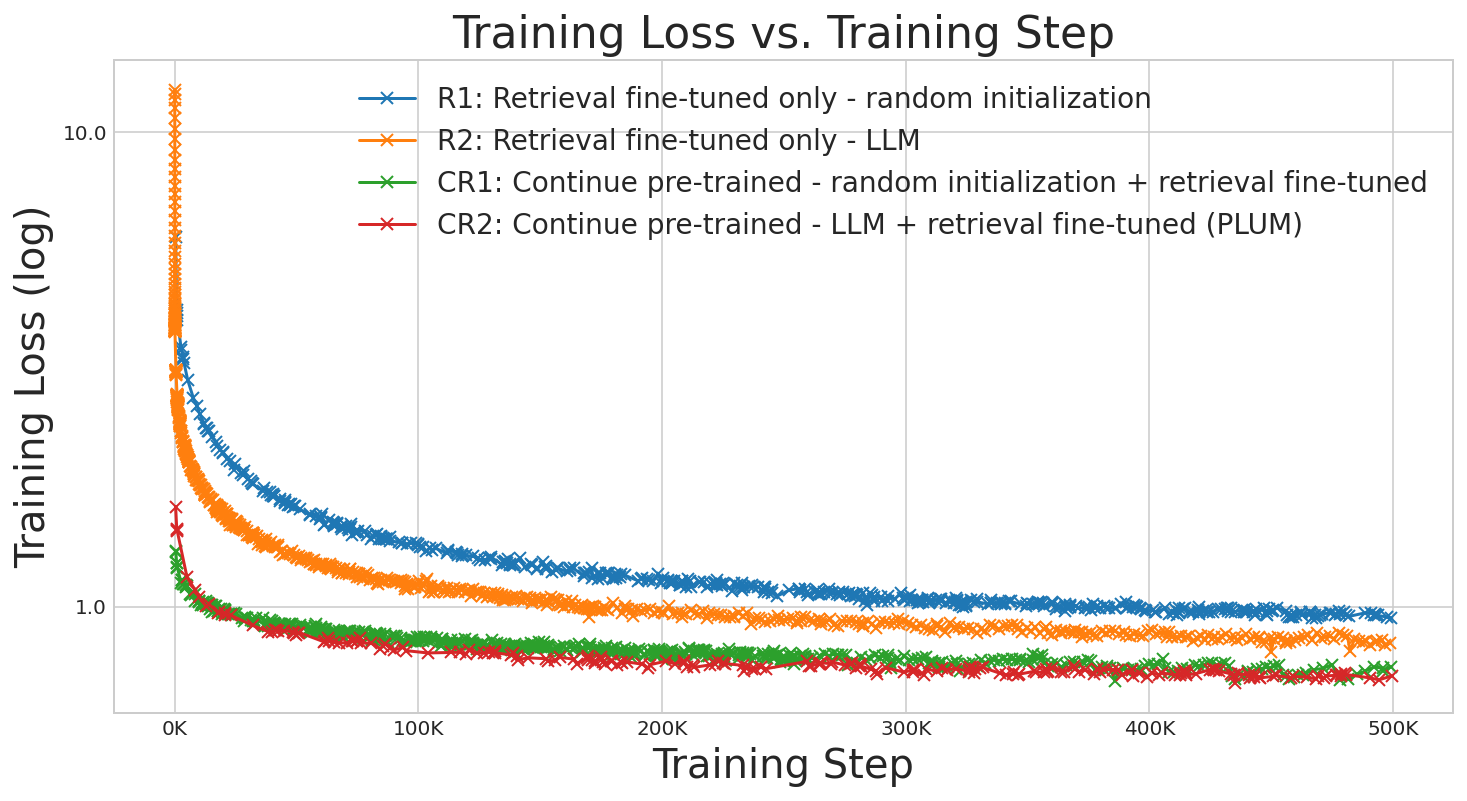}
    \caption{8-th Day Recall@10 and training loss vs retrieval SFT training step.}
    \label{fig:cpt-ablation}
\end{figure*}

\paragraph{Impact of Continued Pre-training.}
The main results are summarized in Table \ref{tab:cpt-ablation}. Following the scaling study, the performance of each of the four models is evaluated on a large, held-out next day dataset, and we report recall@10 as the retrieval metric. There is a large performance gap between R1 and CR1 (or R2 and CR2), showing the benefit of CPT for retrieval fine-tuning. Furthermore, Figure \ref{fig:cpt-ablation} shows the 8-th day recall@10 and the training loss against training steps. It's clear to see that the models with CPT can converge much faster. Therefore, CPT can benefit the training efficiency of task-specific fine-tuning, especially when multiple downstream tasks share the same CPT model.

\begin{table}[h!]
\small
\centering
\caption{Main generative retrieval performance of the four model configurations.}
\label{tab:cpt-ablation}
\begin{tabularx}{0.9\linewidth}{X X X X X}
\toprule
\textbf{Model} & \textbf{Pre-trained LLM} & \textbf{CPT} & \textbf{Recall@10 (8th-day)} \\
\midrule
R1 & No & No  & 0.19 \\
\midrule
R2 & Yes & No  & 0.23 \\
\midrule
CR1 & No & Yes & 0.27 \\
\midrule
CR2 & Yes & Yes & 0.28 \\
\bottomrule
\end{tabularx}
\end{table}

\paragraph{Impact of Pre-trained LLM}
Now we look into the benefits of training from a general-purpose LLM versus training from a randomly initialized state. A consistent pattern emerges from the results in Table \ref{tab:cpt-ablation}: models initialized from a pre-trained LLM consistently outperform their random initialized counterparts with or without CPT. We hypothesize that this advantage stems primarily from the pre-trained LLM’s pre-existing understanding of natural language, or the general sequence processing capability learned through massive LLM pre-training is directly useful for recommendation. A further study of this observation is out of this paper's scope.

\subsection{Scaling Study}
\label{subsec:scaling_study}

In this section, we conduct a scaling study for generative retrieval models with YouTube production data. We use four different sizes from the Gemini-1.5 \citep{team2024gemini} Mixture-of-Experts (MoE) model family: MoE-110M, MoE-370M, MoE-900M, and MoE-3B (here the size indicates the number of activated parameters per token) with their total number of parameters spanning from less than 1B to more than 10B. In particular, we aim to answer: (1) How does loss scale with different model size and compute? (2) How does retrieval metric scale with different model size and compute? (3) What is the best model size for different compute budgets?

\subsubsection{Experiment Setup}

\paragraph{Dataset} Our study was conducted on a production dataset sourced from a key YouTube surface recommending what videos to watch next given a current watched video and other contextual signals like watch history (see Figure 4 in \cite{zhe2019rec}). Here the recommender system employs a multi-stage architecture, encompassing candidate generation, pre-ranking, and ranking, among other phases. Our study centers on the candidate generation stage, evaluating retrieval performance against a comprehensive video corpus containing billions of videos.

\paragraph{Training and Evaluation} Our input prompt has the following format:  ``{{watch history}} | {{user features}} | {{context video features}}", where {{watch history}} is a chronological sequence of watches, each represented by the concatenation of SID tokens, and other feature tokens. For this study, input sequence length is fixed at 1,536 tokens, which can cover roughly 100 most recent watches along with other features. SFT labels are selected using the next watch on YouTube employing a downsampling mechanism based on user engagement and satisfaction signals. SID tokens of the next watch are used as the prediction targets. We randomly shuffle 7 continuous days (from July 2025) of data for training. Evaluation is performed on the subsequent day (Day 8). Our experiments were conducted with 1,024 Google’s v6e Tensor Processing Units (TPUs) each with 32GB High Bandwidth Memory (HBM). We run 4 trainers in parallel each taking 256 TPUs.

\paragraph{Hyperparameter Selection} We conducted a pilot study using the MoE-900M model to determine the model's sensitivity to global batch size and learning rates (see Appendix~\ref{sec:scaling_pilot} for details). Based on the learnings from this pilot study, in each case we set the batch size to approximately saturate the HBM since it indicates larger batch sizes are preferable. For learning rate, using $5 \times 10^{-5}$ for MoE-900M as the anchor point, we use slightly larger learning rates for smaller models (and vice versa) based on our best guess. Note that unlike scaling law studies in natural language processing (e.g.,~\cite{hoffmann2022training}), we employ constant learning rates. This approach is preferred for two reasons: (1) continuous training in production models does not involve a specific stopping step number, and (2) constant learning rates allow for perfect and precise interpolation of all intermediate data points throughout the training process. Hyperparameters of different model sizes are summarized in Table~\ref{tab:scaling-configs}. All models are warm-started from their corresponding Continued Pre-training checkpoints (all with Iso-FLOPS $1 \times 10^{22}$). Continued Pre-training training FLOPs are excluded from all FLOPs numbers in this section.


\begin{table}[ht]
\centering
\caption{Scaling Law Study Configurations.}
\label{tab:scaling-configs}
\begin{tabular}{l c c c}
\toprule
\textbf{Model Size} & \textbf{Learning Rate} & \textbf{Global Batch Size} \\
\midrule
MoE-110M & $1 \times 10^{-4}$ & 25,600  \\
MoE-370M & $7 \times 10^{-5}$ & 15,360  \\
MoE-900M & $5 \times 10^{-5}$ & 7,680 \\
MoE-3B   & $2 \times 10^{-5}$ & 3,584 \\
\bottomrule
\end{tabular}
\end{table}

\subsubsection{Experiment Results}

\begin{figure}[ht]
    \centering
    \begin{subfigure}[t]{0.49\linewidth}
        \centering
        \includegraphics[width=\linewidth]{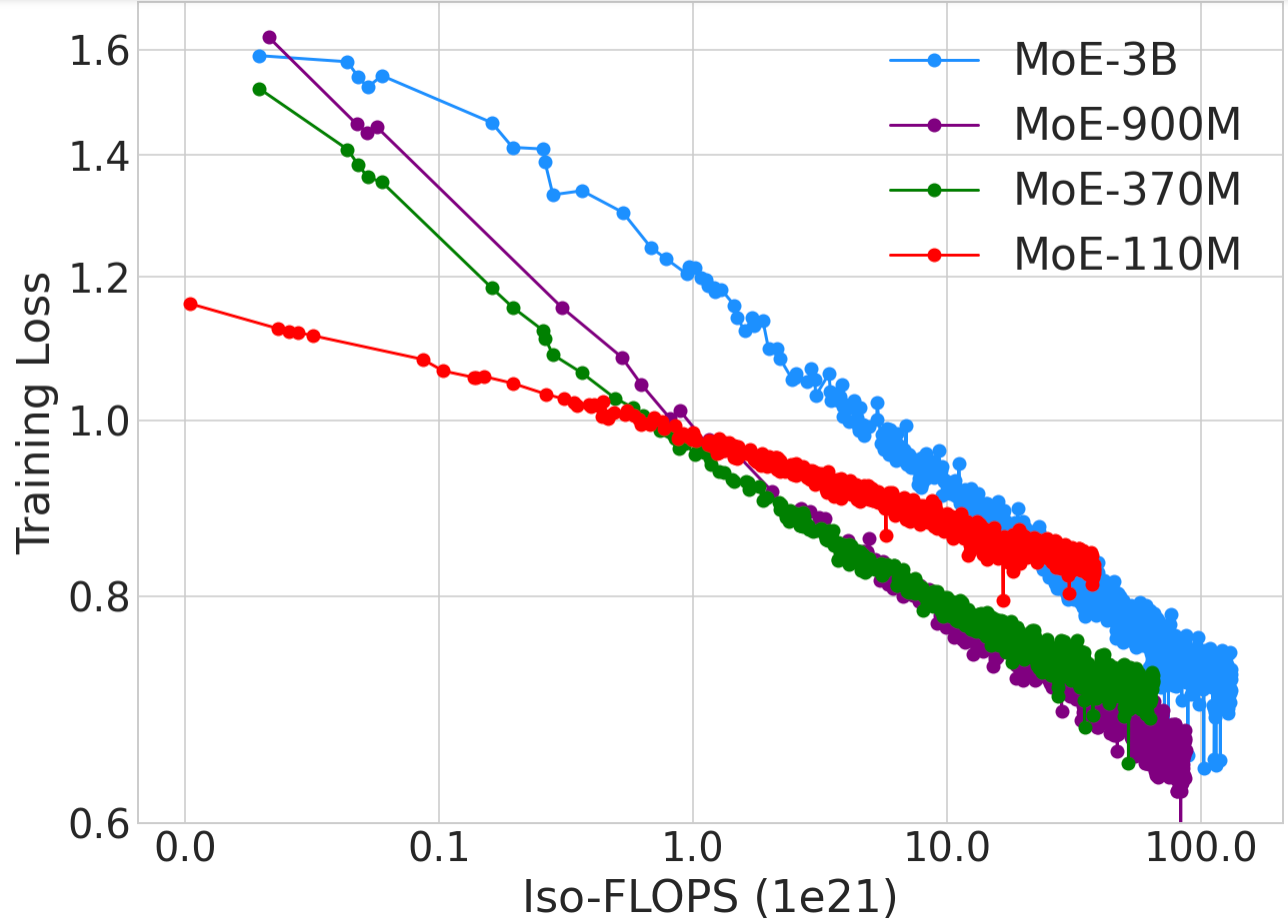}
        \caption{Iso-FLOPS v.s. training loss.}
        \label{fig:scaling-train-loss}
    \end{subfigure}
    \hfill 
    \begin{subfigure}[t]{0.49\linewidth}
        \centering
        \includegraphics[width=\linewidth]{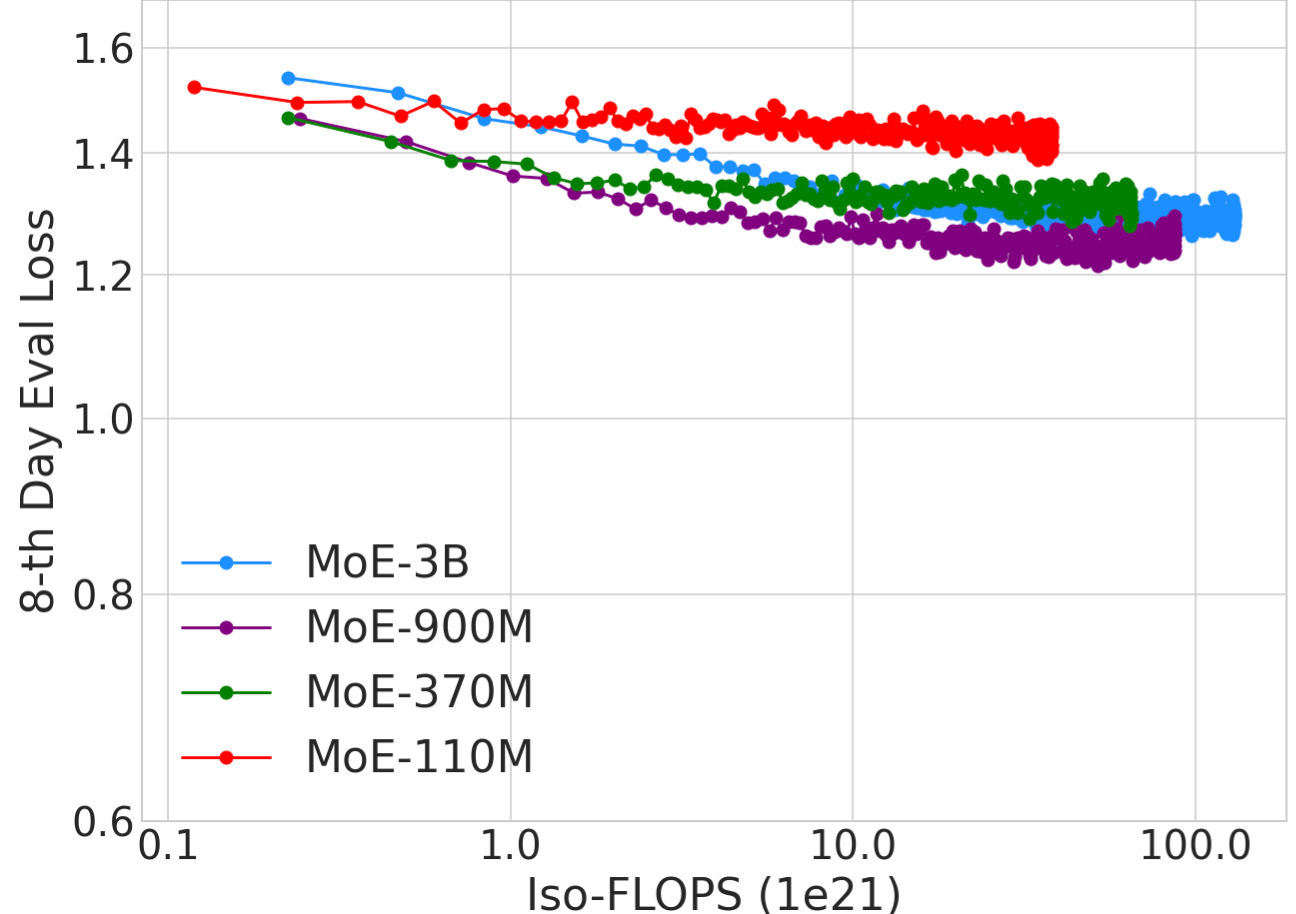}
        \caption{Iso-FLOPS v.s. eval loss.}
        \label{fig:scaling-eval-loss}
    \end{subfigure}
    
    \caption{Training and evaluation loss variation as we scale up training Iso-FLOPS.}
    \label{fig:scaling-loss}
\end{figure}

\paragraph{Loss Scaling with Compute and Model Size} Figure~\ref{fig:scaling-loss} displays the correlation between training/evaluation loss (on day 8) and training Iso-FLOPS. Figure~\ref{fig:scaling-train-loss} reveals a distinct power-law correlation between Iso-FLOPS and training loss for each model size. As Iso-FLOPS increase, the optimal model size, representing the training loss frontier, progressively shifts. This shift begins with MoE-110M (red line), moves to MoE-370M (green line), then to MoE-900M (purple line), and is hypothesized to eventually reach MoE-3B (blue line). Figure~\ref{fig:scaling-eval-loss} illustrates that there is also a power-law correlation between Iso-FLOPS and evaluation loss for each model size before it finally begins to saturate, which we attribute to our extensive training. We also observe that the evaluation loss frontier shifts toward larger models significantly sooner than the training loss, suggesting that larger models exhibit superior generalization to future data distributions.

\begin{figure}[ht]
    \centering
    \begin{subfigure}[t]{0.49\linewidth}
        \centering
        \includegraphics[width=\linewidth]{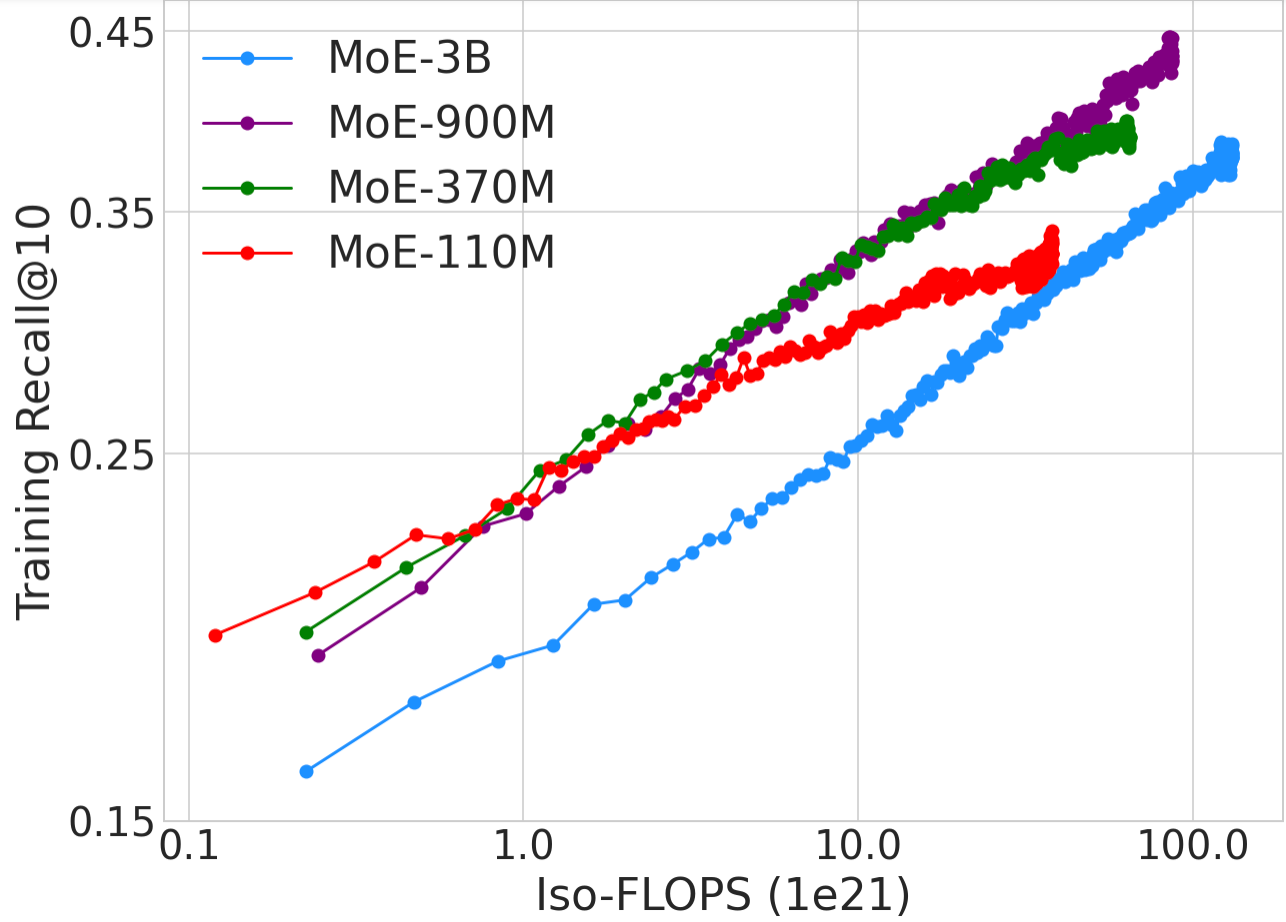}
        \caption{Iso-FLOPS v.s. training Recall@10.}
        \label{fig:scaling-train-recall}
    \end{subfigure}
    \hfill
    \begin{subfigure}[t]{0.49\linewidth}
        \centering
        \includegraphics[width=\linewidth]{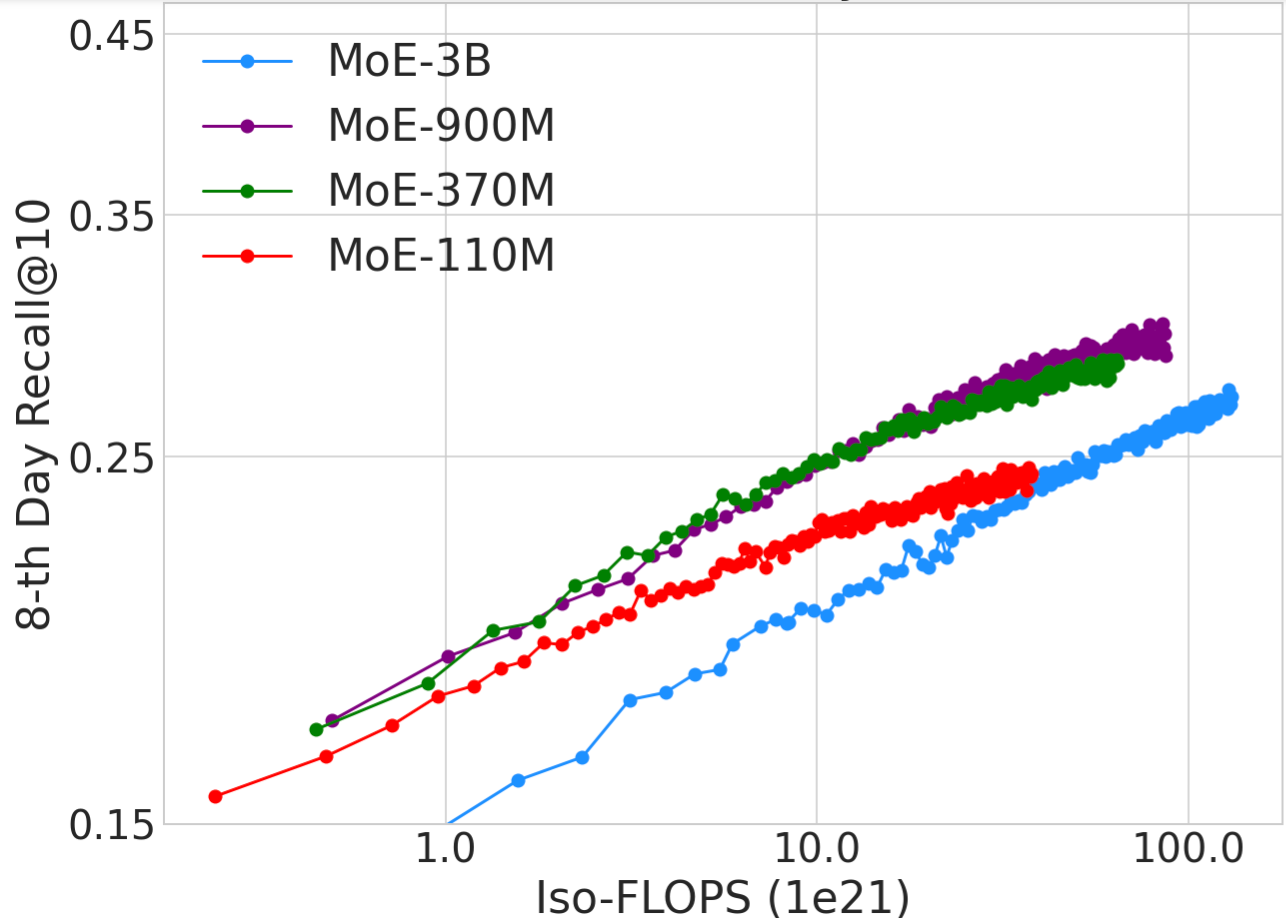}
        \caption{Iso-FLOPS v.s. eval Recall@10.}
        \label{fig:scaling-eval-recall}
    \end{subfigure}
    
    \caption{Training and evaluation Recall@10 variation as we scale up training Iso-FLOPS.}
    \label{fig:scaling-recall}
\end{figure}

\paragraph{Retrieval Metric Scaling with Compute and Model Size} Similarly, in Figure~\ref{fig:scaling-recall} we report the relationship between training/evaluation Recall@10 (on day 8) and training Iso-FLOPS, with both axes presented on a logarithmic scale. The findings are similar to those in Figure~\ref{fig:scaling-loss}, except that in Figure~\ref{fig:scaling-recall} the evaluation Recall@10 shows fewer signs of slowing down to increase, suggesting that model keeps improving the generation of accurate and complete sequences that lead to correct document. 
Additionally, the three smaller models have processed more than one epoch of data, with MoE-110M trained as many as 4.24 epochs but still showing no signs of overfitting (see Figure~\ref{fig:scaling-eval-recall}).


\begin{figure}[ht]
    \centering

    \begin{subfigure}[t]{0.39\linewidth} 
        \centering
        \includegraphics[width=0.95\linewidth]{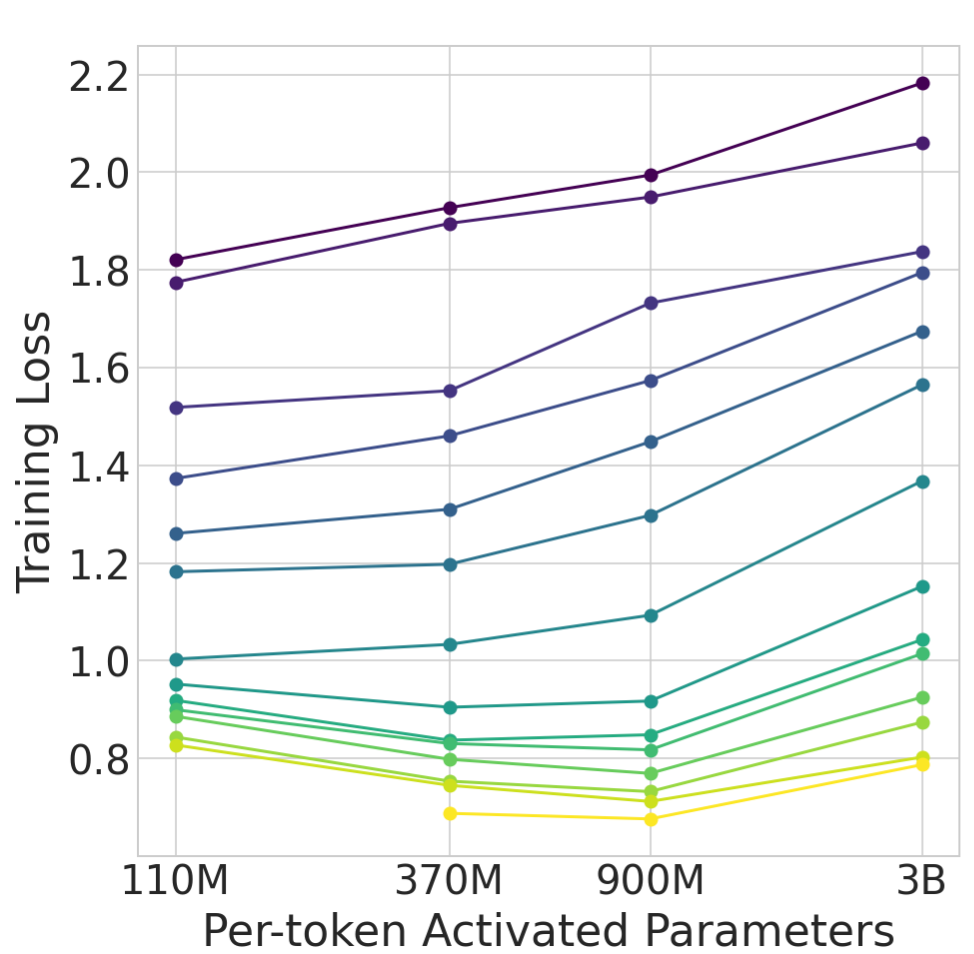} 
        \caption{Training loss for Fixed Compute Budget.}
        \label{fig:scaling-budget-loss}
    \end{subfigure}
    \hfill 
    \begin{subfigure}[t]{0.59\linewidth} 
        \centering
        \includegraphics[width=0.95\linewidth]{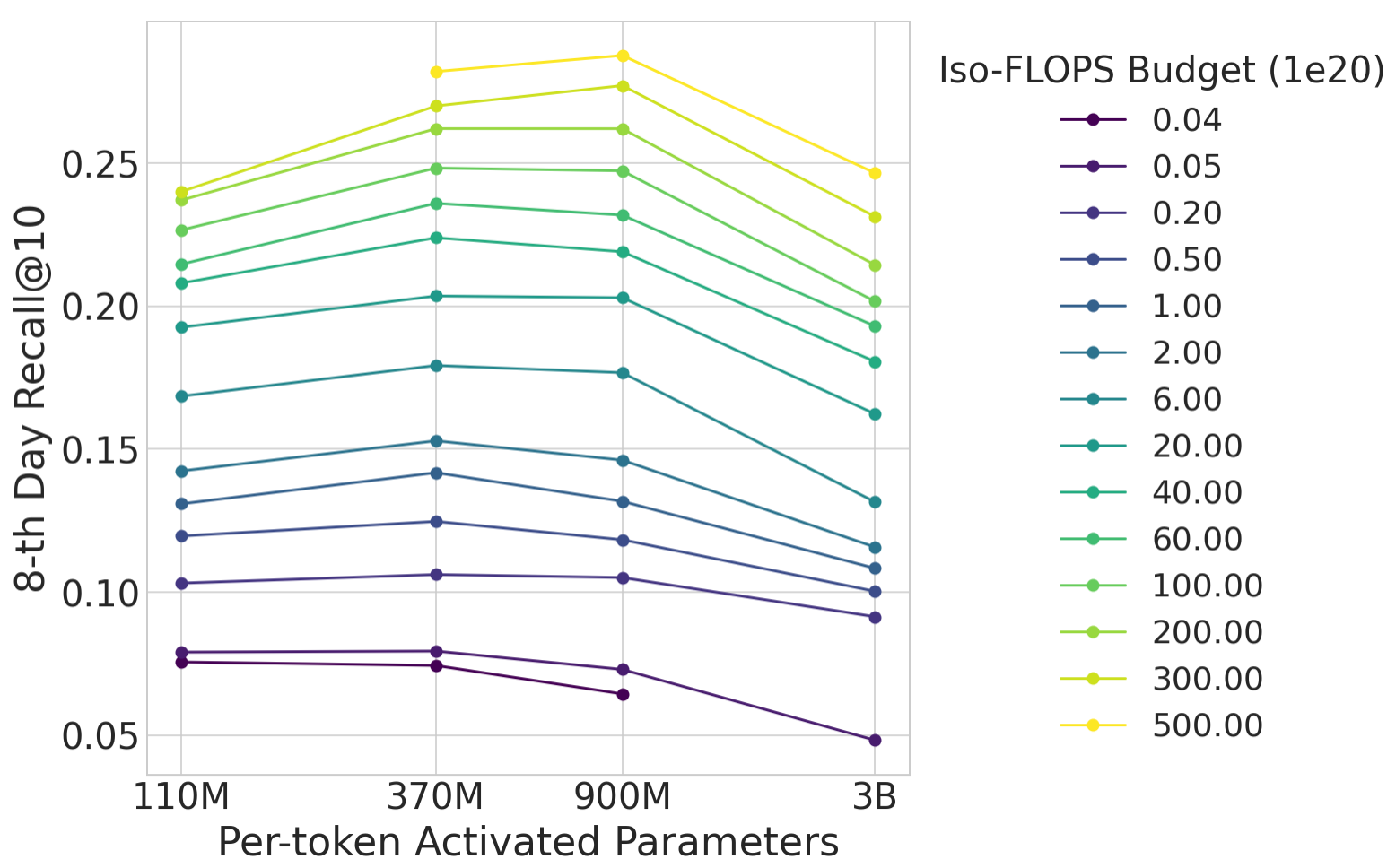}
        \caption{Eval Recall@10 for Fixed Compute Budget.}
        \label{fig:scaling-budget-recall}
    \end{subfigure}
    
    \caption{Iso-FLOPS curves. We evaluate training loss and evaluation Recall@10 at various fixed Iso-FLOPS budgets, selecting the appropriate number of steps for each model size.}
    \label{fig:scaling-budget}
\end{figure}

\paragraph{Optimal Model Size for Fixed Iso-FLOPS Budgets}
In Figure~\ref{fig:scaling-budget}, we demonstrate the gradual shifts of optimal model size to larger ones as we increase the given Iso-FLOPS budgets. Note that our use of constant learning rate for each model enables us to directly assess the model performance at each Iso-FLOPS budget using a single run. Similar to the Chinchilla paper \citep{hoffmann2022training}, in Figure~\ref{fig:scaling-budget-recall} there is a peak for most of the Iso-FLOPS budgets we consider.

\subsubsection{Limitation Discussion}


We note that the MoE-3B model did not outperform the MoE-900M model for the compute budgets we considered. Our scaling study is constrained by compute resource. The suboptimal hyperparameter setup might be a significant contributing factor. For instance, our pilot study indicates a correlation between larger batch sizes and increased training efficiency for the same model size. Since we allocate the same training resources for each model size, using a batch size that saturates the HBMs (see Table~\ref{tab:scaling-configs}), larger models may be at a disadvantage due to their consequently smaller batch sizes. Indeed, at the end of training, the MoE-3B model has only processed 0.57 epochs of training data (about 5B examples), a more than 2x gap from the second largest model. Our scaling study on the generative retrieval task reveals that compute-optimal training requires a synchronized scaling of both training examples and model size.

%% file: conclusion.tex
We introduced PLUM, a framework for adapting pre-trained LLMs for large-scale recommendation tasks. Our method enhances item tokenization with SIDv2, uses a large-scale continued pre-training stage to align the LLM with user behavior and ground SIDs in text, followed by a supervised fine-tuning stage to make the model excel at generative retrieval.


As a launched platform in YouTube, PLUM is able to add unique value to the current system, while being our first neural model built without large embedding tables. Our ablations validate the value of the CPT stage, the benefit of initializing from a pre-trained LLM, and our proposed SID techniques. This paper is a preliminary study on aligning LLMs with real-world recommendation systems, opening future research directions such as applying the PLUM framework to other tasks like ranking and personalized search, developing new decoding strategies for candidate diversity, and enabling seamless generation of SIDs and natural languages.

%% file: ack.tex
We would like to thank the following for their efforts and feedback in the application of the techniques described in this paper to a diverse set of surfaces and use cases (alphabetical order): Saksham Agarwal, Sriraj Badam, Sourabh Bansod, Xilun Chen, Yi Chen, Bernardo Cunha, Onkar Dalal, Mingyan Gao, Elad Ganmor, Danfeng Guo, Pooja Gupta, Ralf Gutsche, Yuan Hao, Chuan He, Lily He, Yan Huang, Kate Jones, Jiewen Lang, Pengfei Li, Wen Li, Li Wei, Sean Liew, Qiguang Liu, Sunny Liu, Yang Liu, Jianan Lu, Yilong Luan, Shifan Mao, Tomer Margolin, Radwa Metwali, Aniruddh Nath, Hardik Patel, Amy Pu, Vasin Punyakanok, Murphy Ren, Yuji Roh, Yuan Shao, Ajay Shekhawat, Fabio Soldo, Yanwei Song, Qian Sun, Jiaxi Tang, Aahan Tyagi, Diego Uribe, Jacky Wang, Ting Wang, Siqi Wu, Bo Yan, Shuhao Ye, Likang Yin, Qiao Zhang, Rein Zhang, Sai Zhang, Vivian Zhang, Yaping Zhang, Gwendolyn Zhao, Zelong Zhao, Zhile Zou, Dav Zimak. We also thank Mahesh Sathiamoorthy and Anima Singh for their profound contributions to the conceptualization and initial development of this project.

%% file: appendix.tex
\subsection{Scaling Pilot Study} \label{sec:scaling_pilot}

As a precursor to our scaling study, we conducted a pilot study using the MoE-900M model. The goal of this pilot is to determine the model’s sensitivity to two hyperparameters: global batch size and learning rates. The outcomes of this pilot will inform the hyperparameter selection for the subsequent, more extensive scaling analysis, ensuring computational resources are utilized effectively.

\subsubsection{Experiment Setup}
We ran four trainers each with 256 TPUs and initialized from the (same) pre-trained checkpoint. The pilot ran for approximately two weeks, a period sufficient to establish clear performance trends. The ``base setup" (Trainer A) uses a constant learning rate of $1 \times 10^{-4}$ and a batch size that saturates the available HBM. The other configurations systematically vary one of these two hyperparameters.

\begin{table}[ht]
\centering
\caption{Pilot Study Configurations with MoE-900M.}
\label{tab:pilot-study}
\begin{tabular}{lcl}
\toprule
\textbf{Trainer} & \textbf{Learning Rate} & \textbf{Batch Size} \\
\midrule
A (Base) & $1 \times 10^{-4}$ & Saturate HBM \\
B        & $1 \times 10^{-4}$ & 0.5x Saturate HBM \\
C        & $2 \times 10^{-4}$ & Saturate HBM \\
D        & $5 \times 10^{-5}$ & Saturate HBM \\
\bottomrule
\end{tabular}
\end{table}

\subsubsection{Results}

\paragraph{Learning Rate Sensitivity} The results indicate that the training process is tolerant of different learning rates within a reasonable range. Specifically, the performance difference between the base learning rate of $1 \times 10^{-4}$ (Trainer A) and the lower rate of $5 \times 10^{-5}$ (Trainer D) was not statistically significant within the duration of the pilot study. However, doubling the learning rate to $2 \times 10^{-4}$ (Trainer C) resulted in substantially worse performance, indicating that this rate is likely outside the optimal range for this model and dataset.

\paragraph{Batch Size Sensitivity} Using a smaller batch size harms training efficiency. Halving the batch size requires more than double the training steps to achieve a comparable level of performance. This highlights the importance of maximizing batch size to the limits of hardware memory for compute optimal training.

\subsection{In-context Learning Examples} \label{app:in-context-learning}

Table \ref{tab:fewshot_topic_gen} shows a few examples to demonstrate the in-context learning capability of the model after CPT with Semantic IDs. The CPT model based on pre-trained LLM was able to correctly complete the sentence with a semantically appropriate phrase, by following the few-shot task description. In contrast, if we apply the same recs pre-training on a randomly-initialized model, the model struggled to form a coherent phrase and made distinction between SID and text tokens. 

\begin{table*}[t]
    \centering
    \footnotesize 

    \begin{minipage}{\linewidth}
        \centering
        \subcaption{Example 1}
        \label{subtab:topic_gen}
        \begin{tabular}{p{0.3\linewidth} p{0.65\linewidth}} 
            \toprule
            \textbf{Video Thumbnail} & \includegraphics[width=1.0\linewidth]{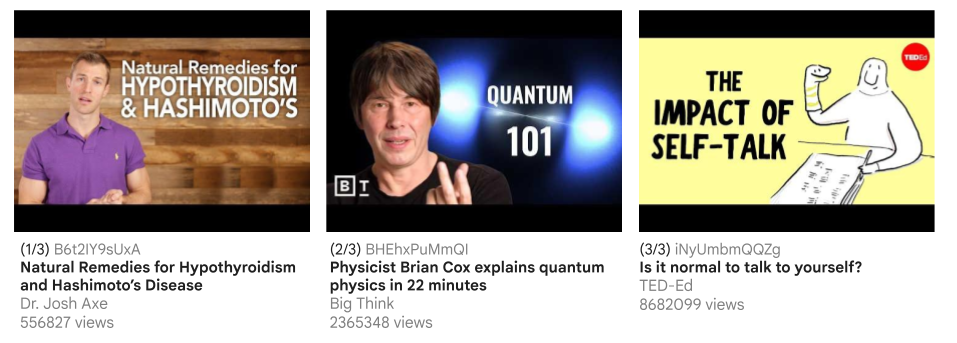} \\
            \midrule
            \textbf{Few-shot Input} & The video \texttt{A1991} ... \texttt{H10} is about health and pain management. \\
             & The video \texttt{A364} ... \texttt{H37} is about the basics of quantum physics. \\
             & The video \texttt{A37} ... \texttt{H25} is about \\
            \midrule
            \textbf{LLM-Initialized CPT Output} & psychology and mind \\
            \midrule
            \textbf{Randomly-Initialized CPT Output} & 100\% video \texttt{A252 H7} \\
            \bottomrule
        \end{tabular}
    \end{minipage}
    
    \vspace{4mm}
    
    \begin{minipage}{\linewidth}
        \centering
        \subcaption{Example 2}
        \label{subtab:learning_examples}
        \begin{tabular}{p{0.3\linewidth} p{0.65\linewidth}} 
            \toprule
            \textbf{Video Thumbnail} & \includegraphics[width=1.0\linewidth]{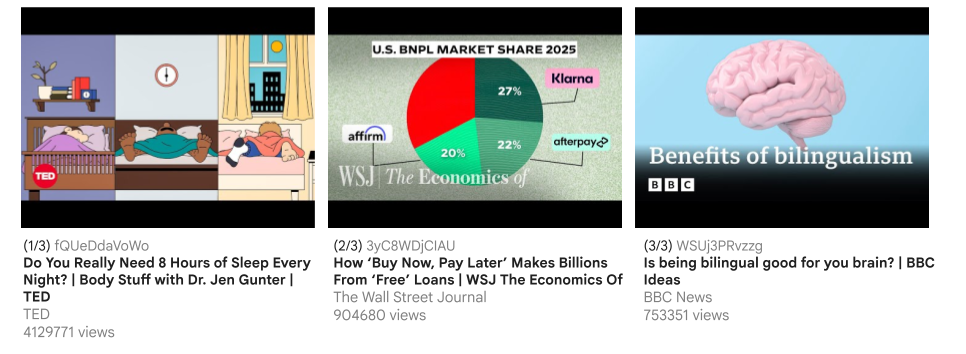} \\
            \midrule
            \textbf{Few-shot Input} & Video \texttt{A37} ... \texttt{H41} answers the question: do you really need 8 hours of sleep? \\
            & Video \texttt{A1926} ... \texttt{H8} answers the question: how buy-now pay-later loan business makes profit? \\
            & Video \texttt{A1882} ... \texttt{H32} answers the question: \\
            \midrule
            \textbf{LLM-Initialized CPT Output} & the impact of language in your life. \\
            \midrule
            \textbf{Randomly-Initialized CPT Output} & - - - 100\% -1999-11-16 \\
            \bottomrule
        \end{tabular}
    \end{minipage}

    \caption{Qualitative examples of few-shot, in-context learning example. Each table shows the text-based prompt and model's output from two different initializations. The thumbnails (top) show the visual content for each corresponding Semantic ID in the prompt.}
    \label{tab:fewshot_topic_gen}
\end{table*}